\documentclass[11pt]{article}
\usepackage{amsmath,amsthm,amssymb}
\usepackage[a4paper]{geometry}
\usepackage{graphicx}
\usepackage[round]{natbib}
\usepackage{float}
\usepackage{subcaption}
\usepackage{url}
\usepackage{multirow}
\usepackage[hidelinks,colorlinks=true,linkcolor=blue,citecolor=blue,urlcolor=blue]{hyperref} 
\usepackage{placeins}
\usepackage{afterpage}
\usepackage{titling}
\makeatletter
\newcommand*\if@single[3]{%
  \setbox0\hbox{${\mathaccent"0362{#1}}^H$}%
  \setbox2\hbox{${\mathaccent"0362{\kern0pt#1}}^H$}%
  \ifdim\ht0=\ht2 #3\else #2\fi
  }
\newcommand*\rel@kern[1]{\kern#1\dimexpr\macc@kerna}
\newcommand*\widebar[1]{\@ifnextchar^{{\wide@bar{#1}{0}}}{\wide@bar{#1}{1}}}
\newcommand*\wide@bar[2]{\if@single{#1}{\wide@bar@{#1}{#2}{1}}{\wide@bar@{#1}{#2}{2}}}
\newcommand*\wide@bar@[3]{%
  \begingroup
  \def\mathaccent##1##2{%
    \if#32 \let\macc@nucleus\first@char \fi
    \setbox\z@\hbox{$\macc@style{\macc@nucleus}_{}$}%
    \setbox\tw@\hbox{$\macc@style{\macc@nucleus}{}_{}$}%
    \dimen@\wd\tw@
    \advance\dimen@-\wd\z@
    \divide\dimen@ 3
    \@tempdima\wd\tw@
    \advance\@tempdima-\scriptspace
    \divide\@tempdima 10
    \advance\dimen@-\@tempdima
    \ifdim\dimen@>\z@ \dimen@0pt\fi
    \rel@kern{0.6}\kern-\dimen@
    \if#31
      \overline{\rel@kern{-0.6}\kern\dimen@\macc@nucleus\rel@kern{0.4}\kern\dimen@}%
      \advance\dimen@0.4\dimexpr\macc@kerna
      \let\final@kern#2%
      \ifdim\dimen@<\z@ \let\final@kern1\fi
      \if\final@kern1 \kern-\dimen@\fi
    \else
      \overline{\rel@kern{-0.6}\kern\dimen@#1}%
    \fi
  }%
  \macc@depth\@ne
  \let\math@bgroup\@empty \let\math@egroup\macc@set@skewchar
  \mathsurround\z@ \frozen@everymath{\mathgroup\macc@group\relax}%
  \macc@set@skewchar\relax
  \let\mathaccentV\macc@nested@a
  \if#31
    \macc@nested@a\relax111{#1}%
  \else
    \def\gobble@till@marker##1\endmarker{}%
    \futurelet\first@char\gobble@till@marker#1\endmarker
    \ifcat\noexpand\first@char A\else
      \def\first@char{}%
    \fi
    \macc@nested@a\relax111{\first@char}%
  \fi
  \endgroup
}
\makeatother

\theoremstyle{plain}
\theoremstyle{definition}

\newtheorem{rema}{Remark}

\newcommand{\EE}[2][]{\mathbb{E}_{#1}\left[#2\right]}

\newcommand{\argmin}{\operatorname{argmin}}
\newcommand{\PP}[2][]{\mathbb{P}_{#1}\left[#2\right]}

\newcommand\indep{\protect\mathpalette{\protect\independenT}{\perp}}
\def\independenT#1#2{\mathrel{\rlap{$#1#2$}\mkern2mu{#1#2}}}
\begin{document}
\title{Estimating Treatment Effect Heterogeneity in Psychiatry:\\ A Review and Tutorial with Causal Forests}

\author{
Erik Sverdrup\thanks{
Department of Econometrics \& Business Statistics, Monash University, Melbourne, Australia.
}
\and 
Maria Petukhova\thanks{
\noindent Department of Health Care Policy, Harvard Medical School, Boston, Massachusetts.
}
\and
Stefan Wager\thanks{
Graduate School of Business, Stanford University, Stanford, California.
\newline Corresponding Author: Erik Sverdrup, Department of Econometrics \& Business Statistics, Monash University, 29 Ancora Imparo Way, Clayton, VIC 3168, Australia; Telephone: +61 990 55 041; E-Mail: \url{erik.sverdrup@monash.edu}.
}
}

\thanksmarkseries{arabic}
% \date{}
\maketitle

\begin{abstract}
Flexible machine learning tools are increasingly used to estimate heterogeneous treatment effects. This paper gives an accessible tutorial demonstrating the use of the \emph{causal forest} algorithm, available in the \texttt{R} package \texttt{grf}. We start with a brief non-technical overview of treatment effect estimation methods, focusing on estimation in observational studies; the same techniques can also be applied in experimental studies. We then discuss the logic of estimating heterogeneous effects using the extension of the random forest algorithm implemented in \texttt{grf}. Finally, we illustrate causal forest by conducting a secondary analysis on the extent to which individual differences in resilience to high combat stress can be measured among US Army soldiers deploying to Afghanistan based on information about these soldiers available prior to deployment. We illustrate simple and interpretable exercises for model selection and evaluation, including targeting operator characteristics curves, Qini curves, area-under-the-curve summaries, and best linear projections. A replication script with simulated data is available at \href{https://github.com/grf-labs/grf/tree/master/experiments/ijmpr}{github.com/grf-labs/grf/tree/master/experiments/ijmpr}. 
\end{abstract}


\section{Introduction}\label{sec:intro}
An important question in psychiatry as well as other scientific disciplines that evaluate effects of interventions on the health and well-being of individuals is the extent to which the treatment effects estimated in typical experimental evaluations differ across individuals \citep{angus2021heterogeneity}. This is an important question because we have long known that the effects of interventions may vary across different segments of the population \citep{kravitz2004evidence}. Because of these differences, some interventions that are estimated not to be effective in the entire population based on the typical assumption of constant treatment effects are nonetheless effective in some important segments of the population, whereas some interventions found to be effective in the entire population are either not effective or in some cases even harmful in important segments of the population \citep{kent2016risk}. In addition, when multiple interventions are available, comparative effectiveness can differ across individuals and population segments \citep{velentgas2013developing}. 

Research on heterogeneity of treatment effects (HTE), as this variation in intervention effects has come to be known, has proliferated in recent years in psychiatry \citep[e.g.,][]{feczko2019heterogeneity, allsopp2019heterogeneity, kaiser2022heterogeneity, cohen2018treatment}, and many other disciplines. The basic approach is to search for significant variations in treatment effects across subsamples. This is most easily done by estimating interactions between some hypothesized effect modifiers assessed prior to randomization and random assignment. When the interaction is non-zero, HTE is said to exist with respect to the specified variable---meaning that the treatment effect differs depending on the value of the variable.

Numerous small clinical trials have reported HTE detections of this sort with respect to such fundamental baseline variables as patient age, sex, education, and symptom severity \citep[e.g.,][]{cuijpers2014gender, driessen2010does}. But a question arises in trying to synthesize all this information when numerous significant interactions of this sort emerge, as meta-analyses show that they do \citep[e.g.,][]{maj2020clinical}. What is the best way to combine the information about all these specifiers into a single composite? A commonly used method is to apply the potential outcomes framework \citep{imbens2015causal} in analyzing the results of a moderated multiple regression model that includes multiple interactions. Counterfactual logic is used here to generate two predicted outcome scores from this model for each participant in the study, the first assuming that the participant was assigned to the intervention group and the second assuming that the participant was assigned to the control group \citep{derubeis2014personalized}. The two within-person predicted outcome scores are then compared at the individual level and used as an estimate of the treatment effect for an individual based on the multivariate specifier profile of the individual. The distribution of these differences scores is then examined to investigate the pattern of significant individual differences in the treatment effect. 

A drawback of this approach is that in trying out many different regression models, applications of the procedure that do not account for multiple testing may lead to false positives \citep{van2019models}. This problem can be addressed, though, using data-driven machine learning methods to search for interactions with cross-validation to address the problem of over-fitting \citep{vanderweele2019selecting, wager2018estimation}. The use of data-driven methods in this way also addresses the problem that the moderated regression approach only allows for a limited range of typically two-way interactions, whereas HTE can involve much more complex forms of interaction. Machine learning methods, in comparison, allow for these more complex forms of interaction to be discovered using non-parametric methods. A large body of work has made tremendous progress in adapting a variety of machine learning algorithms to estimate HTE. Examples include \citet{athey2016recursive, wager2019grf, hahn2020bayesian, kennedy2023towards, kunzel2019metalearners, luedtke2016super, nie2020quasi, tian2014simple}.

\section{Estimating treatment effects in non-experimental settings} \label{sec:ate}
To communicate statements about causal effects it is helpful to employ a universal language with well-defined constructs. We employ the potential outcomes framework \citep{imbens2015causal} where we posit the existence of potential outcomes
$Y_i(0)~\text{and}~ Y_i(1)$
which records the hypothetical clinical outcome $Y_i$ for patient $i$ in either treatment state. We typically denote the control state by $W_i = 0$ and the treatment state by $W_i = 1$. Naturally, we only observe one potential outcome for any single patient: the outcome corresponding to the treatment arm she was assigned to. Table \ref{tab:POs} shows an example of hypothetical potential outcomes along with realized outcomes.
\begin{table}[ht]
\footnotesize
\centering
\begin{tabular}{cclll}
Patient $i$ & Treatment assignment $W_i$ & $Y_i(1)$ & $Y_i(0)$ & Realized outcome $Y_i$ \\
\hline
1      & 1 (Treated) & 0 (Sick)         & 0 (Sick) & 0 (Sick) \\
2      & 0 (Control) & 1 (Healthy)  & 0 (Sick)    & 0 (Sick) \\
\vdots & 1 (Treated) & 1 (Healthy)     & 0 (Sick)    & 1 (Healthy)          
\end{tabular}
\caption{Potential outcomes denoting hypothetical outcomes for each patient and each intervention.}
\label{tab:POs}
\end{table}

The assumption that our column of realized outcomes is consistent with the columns of potential outcomes, $Y_i=Y_i(W_i)$, is called the Stable Unit Treatment Value Assumption (SUTVA). In some complicated social network settings, this assumption may not be viable. Consider, for example, if patient A is given behavioral therapy and patient B is not. If patient A interacts with patient B through some social functioning, like a support group, then A's newly acquired behavioral knowledge may influence B. Here, however, we rely on SUTVA and thus rule out scenarios like these by assuming no interference or spillover between patients.

The type of causal effects we are interested in quantifying are average differences in potential outcomes. Given some hypothetical target population, the average treatment effect (ATE) is
$$
\tau = \EE{Y_i(1) - Y_i(0)},
$$
where $\EE{\cdot}$ denotes an average taken over a population. This would quantify, following the setup in Table \ref{tab:POs}, for example, the increase in healthy prevalence when assigning every patient the intervention, as opposed to withholding the intervention from everyone.

Even though for any single patient $i$ we never observe the difference $Y_i(1) - Y_i(0)$, we still have enough information to identify the average difference. If we collect a representative sample from our population and then randomly assign an intervention to some units in the sample while withholding it from others, then the simple difference in means (i.e., the prevalence among treated patients minus the mean prevalence among control patients) provides an unbiased estimate of the average treatment effect. The fact that the treatment assignment is randomized means that our hypothetical potential outcomes and treatment status are decoupled, which justifies this simple calculation.

The availability of such a randomized treatment assignment is typically constrained to randomized controlled trials. In many important settings, due to ethical and resource considerations, we don’t have access to a randomized intervention. A dataset where the intervention assignment is not randomized is often referred to as an observational dataset. As an example, via electronic health records, we could assemble a dataset that records an intervention, such as psychiatric hospitalization, together with an outcome, such as subsequent suicide, in a sample of patients presenting at an emergency department with suicidality. The decision whether to hospitalize such patients is clearly not randomized but will rather depend on attending physician and hospital protocols, patient severity and perceived risk of suicidal behavior, available outpatient alternatives to hospitalization (e.g.,  day treatment programs), and availability of hospital beds. Some of these determinants of hospitalization are also likely to be determinants of subsequent suicidal behavior, which means that a simple comparison of the rates of suicidal behavior in, say, six months after emergency department presentation cannot be interpreted as providing evidence of the effects of hospitalization on these behaviors. At the same time, constructing a randomized controlled trial in a setting of this sort poses ethical challenges.

We can, however, under suitable circumstances, make causal inferences from such data if we are willing to make certain assumptions about access to information about the determinants of treatment assignment that are also determinants of subsequent suicidal behavior in the absence of treatment assignment. The key assumption in this regard is known as the unconfoundedness assumption \citep{rosenbaum1983central}, which posits that if we account for a set of observed patient characteristics $X_i$ known prior to intervention, then we can think of the treatment assignment as if it were randomized,
$$
\left(Y_i(1),~ Y_i(0) \indep W_i\right) \mid X_i ~~ \text{(unconfoundedness)}.
$$
An important object under this setup is the propensity score,
$$
e(x) = \EE{W_i \mid X_i=x},
$$
which is the probability that a patient with confounder variables equal to $x$ is assigned the treatment. To estimate causal effects, a necessary condition is that the association between baseline predictors and intervention assignment is not so strong that no individuals have a non-zero probability of assignment to any of the interventions under study. This is the overlap assumption, which requires that the propensity score $e(x)$ is strictly above 0 and below 1 for every unit $x$. For example, if 100\% of the suicidal patients presenting at an emergency department with severe thought disturbance were hospitalized, it would be impossible to use observational methods to evaluate the effect of hospitalization in preventing subsequent suicidal behaviors among such patients. It is only in the subset of patients whose probability of hospitalization is less than 100\% and greater than 0\% that data-driven estimates of treatment effects can be made. Assessing this requirement empirically translates into ensuring that estimated propensity scores are not close to either zero or one, recognizing that estimation of treatment effects is impossible in the subset of patients whose scores are close to zero or one.

While the overlap assumption is testable, the direct conditional independence assumption under unconfoundedness is not, as it involves statements of independence between random variables. This is usually referred to as an identifying assumption and is fundamentally untestable.
Instead, the evaluation of identifying assumptions requires domain knowledge and subject-matter expertise to assess how credible they are in a particular application. In this tutorial, we are treating this first step as done or known, i.e., the underlying causal graph \citep{pearl1995causal} is agreed upon. In the context of the example of psychiatric hospitalization as a predictor of subsequent suicidal behavior among suicidal emergency department patients, an example where this assumption is invoked is \citet{jamagrf}.

To estimate average treatment effects in observational settings under unconfoundedness, we must account for how the different types of patients respond to and are assigned the treatment before making comparisons between average outcome scores. We can do this by estimating the propensity score. It is also possible to perform the confounding adjustment by estimating conditional outcome functions or via a doubly robust combination of estimates of both the propensity and the conditional outcome function. This is the approach \texttt{grf} implements in the function \texttt{average\_treatment\_effect}; see \citet{ding2023first}, \citet{hernan2023causal}, and \citet{wagerbook} for recent textbook treatments of these methods.

\subsection{From ATE to HTE}

A limitation of average treatment effects is that they can mask important individual differences. In the example of hospitalization for suicidal patients presenting to an emergency department, even though numerous observational studies that attempted to adjust for baseline confounders \citep{steeg2018suicide, carroll2016variation, large2018psychiatric, jones2008treating} and two small experiments \citep{large2018psychiatric, jones2008treating} were all unable to detect a non-zero average treatment effects of hospitalization in preventing subsequent suicidal behavior, a subsequent observational study \citep{jamagrf} documented the presence of HTE by finding a subset of patients with significantly positive average treatment effects (i.e., these patients were helped by hospitalization) and finding a second subset of patients with significantly negative average treatment effects (i.e., these patients were harmed by hospitalization). These subsets were found by focusing on something known as conditional average treatment effects (CATEs).

CATEs are averages of the individual-level treatment effects (ITEs),
$
\Delta_i = Y_i(1) - Y_i(0),
$
which represent individual-level responses to treatment. If known, ITEs would directly represent a measure of HTE. Unfortunately, the ITE is not possible to estimate, given that we only get to observe one realized potential outcome for every patient $i$. An object that in granularity, lies somewhere between the ITE and the ATE is the CATE,
$$
\tau(x) = \EE{Y_i(1) - Y_i(0) \mid X_i = x}.
$$
The CATE is an average treatment effect over patients $i$ that have characteristics that equal $x$, where $x$ is a collection of pre-treatment numeric predictor variables (e.g., age, sex, blood pressure, past trauma, etc.). A certain patient might have predictor variables that look like
$$
x = \{55,~ \text{female},~ 120,~ \text{no},~ \ldots\}.
$$
The CATE is simply an average treatment effect that is conditional on the patient having a multivariate profile defined by this vector of characteristics. (To keep the exposition simple and limit notational sprawl, $X_i$ can refer to both confounders and treatment effect modifiers, although the two sets of variables do not have to be the same).

This more granular object has practical appeal since we can formulate it as a function, denoted by $\tau(\cdot)$, that takes as input some patient characteristics given by $x$, then maps it to a treatment effect, as given by the ATE for the subset of patients with this profile. An individualized treatment rule could be to treat those patients where the treatment effect, as measured by CATE, is sufficiently large. If only $x$ is known, then CATE thresholding is optimal \citep[e.g.,][]{bhattacharya2012inferring}.

\section{The causal forest approach to estimating CATE}\label{sec:CF}

Causal forests \citep{wager2019grf} build on Breiman’s random forest algorithm \citep{breiman2001random}, a popular and empirically successful algorithm for prediction that allows for a mix of real- and discrete-valued predictors. Random forests have a strong empirical track record of doing well out-of-the-box in many applications, often with minimal tweaking of tuning parameters. An appealing aspect of random forests is that they essentially act as an algorithmic way of dividing data into subgroups and then making predictions based on which subgroup a patient belongs to. A core component of the random forest algorithm is to scan over each patient characteristic, then decide on a cut point that serves as a ``good'' candidate for a subgroup partition. The algorithm then aggregates these partitions into neighborhoods where patients with specific characteristics in terms of predictor variables $X_i$ have similar values on outcome $Y_i$. 

At first glance, the algorithmic blueprint of random forests appears like a promising candidate for our task of discovering subgroups with different treatment effects, as we are looking for ways to partition a sample of participants in an intervention experiment according to observable characteristics. The primary difference between causal forests and Breiman's random forests is that the objective of the original random forests is to partition the sample to optimize discrimination of individual differences in scores on an outcome, whereas our objective is to predict differential treatment response.

\begin{figure}[ht]
    \centering
    \begin{subfigure}[b]{0.475\textwidth}  
        \centering 
        \includegraphics[width=\textwidth]{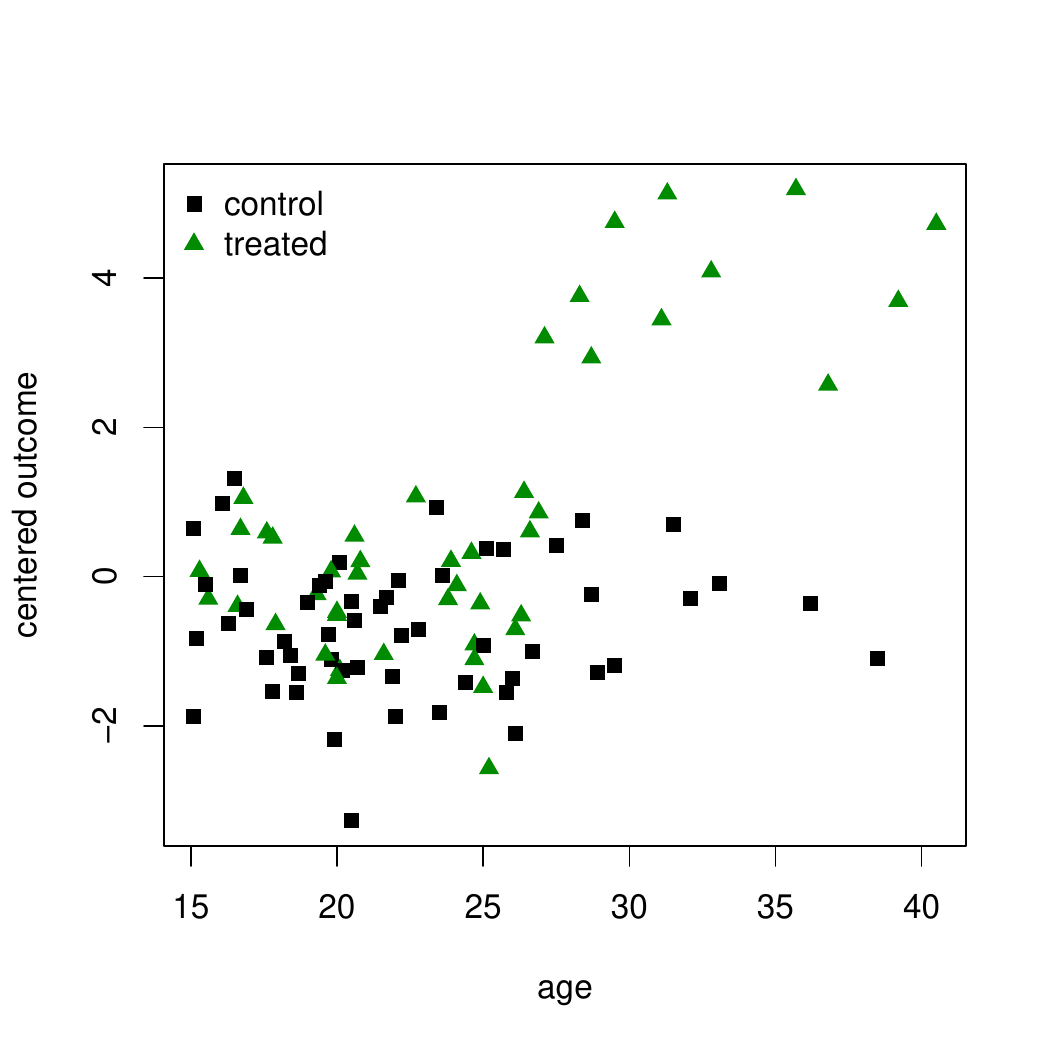}
        \caption[]
        {{Centered outcomes ($\widetilde Y_i$) for control and treatment groups are plotted against $age$, a potential predictor variable.\\\\\\}}    
        \label{fig:cf_fig1}
    \end{subfigure}
    \hfill
    \begin{subfigure}[b]{0.475\textwidth}
        \centering
        \includegraphics[width=\textwidth]{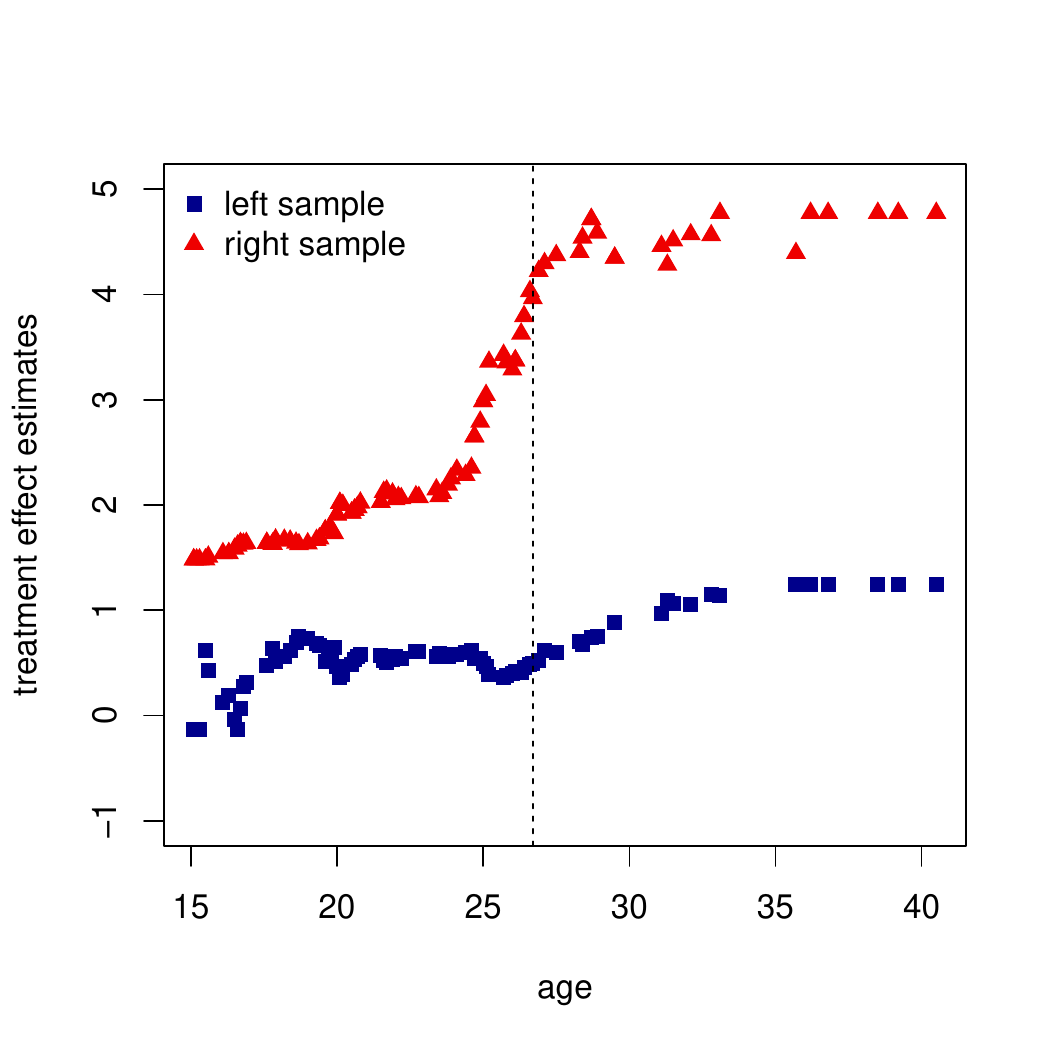}
        \caption[]
        {{Estimated average treatment effects ($\hat \tau_{\text{left}}$ and $\hat \tau_{\text{right}}$) for subgroups split by $age$. The dashed vertical line indicates an optimal split point where the difference in treatment effect estimates between the left (blue) and right (red) samples is largest.}}    
        \label{fig:cf_fig2}
    \end{subfigure}    
    \caption{Stylized example of splitting to maximize treatment effect heterogeneity using centered outcomes and treatments.}
    \label{fig:cf_fig}
\end{figure}
The challenge in targeting treatment heterogeneity that does not arise in the basic prediction setting is that scores on an outcome variable are observed for each participant, whereas individual-level treatment effects are not observed. As we saw previously, though, treatment effect estimation with non-experimental data can be carried out by adjusting for baseline characteristics. We do this using a two-step process. In the first step, causal forests account for uneven treatment assignments via the propensity score $e(x)$, as well as baseline effects by estimating the conditional mean (marginalizing over treatment)
$$
m(x) = \EE{Y_i \mid X_i=x}.
$$
In this first step, baseline effects are essentially stripped away via something known as an orthogonal construction \citep{DML, robinson1988root}. This construction works on the outcomes as well as on treatment indicators to generate residualized outcome $\widetilde Y_i = Y_i - m(X_i)$ and treatment indicators $\widetilde W_i = W_i - e(X_i)$ that have been adjusted for confounding effects. 
In the second step, $\widetilde Y_i$ and $\widetilde W_i$ are used to form treatment effects estimates, and a random forest is used to find predictor variable partitions where these treatment effects estimates exhibit heterogeneity. Figure \ref{fig:cf_fig} presents a stylized illustration of the underlying logic for this phase of the algorithm. For a candidate predictor variable, such as age, we seek a single axis-aligned split separating units into two groups with different average treatment effect estimates. Here, units with ages above 25 appear to benefit more, and a partition that immediately increases heterogeneity, as measured by the vertical distance between estimates in Figure \ref{fig:cf_fig2}, would be at this point. This exercise is repeated recursively for the left and right samples, considering all possible predictor variables (e.g., blood pressure) to partition the covariates $X_i$ into regions where treatment effects are expected to differ.

In the case where the analysis is applied to an experiment, where the probability of assignment to each intervention arm is known by construction, these probabilities can simply be supplied in the first step. However, in a non-experimental situation in which we are attempting to approximate estimates of treatment effects by adjusting for nonrandom exposure respective to observed baseline covariates, the probabilities of treatment assignment are estimated in the first step with separate random forests.

It is possible to modify this random forest framework to accommodate many other relevant statistical quantities and settings -- thus the name “generalized” random forests--\texttt{grf}. For example, \citet{wager2019grf} use this general framework to construct an instrumental forest that estimates HTE in settings with an instrumental variable, \citet{cui2023estimating} use this framework to construct a causal survival forest for estimating HTE with right-censored survival outcomes, and \citet{friedberg2020local} constructs a local linear forest. While these random forest algorithms are designed to directly target various causally relevant targets, we emphasize that one still needs to exercise caution when interpreting flexible non-parametric point estimates of these quantities. To give some intuition, machine learning models serve as great complements to traditional parametric statistical models as they allow us to be model-agnostic. This agnostic property, however, comes at the cost of introducing higher estimation uncertainty. So, our estimates of individual-level treatment effects come with considerable uncertainty, and in assessing the quality of these estimates, it is practical to move to a less granular summary measure. In the applied Section \ref{sec:application}, we cover different approaches to assess if the individual-level estimates capture meaningful heterogeneity via calibration exercises. On a general note, the perspective we are taking is that machine learning tools can serve as effective algorithmic devices that target CATEs, which an analyst can then validate on held-out data and transparently convey potentially interesting scientific findings regarding treatment heterogeneity.

Finally, it is worth pointing out that the underlying identifying causal assumptions used in causal forests are the same as in classical parametric models, such as in the moderated regression approach described earlier. However, a main difference is that whereas classical parametric models (e.g., linear and logistic regressions) specify particular predictors and functional forms of their associations with the outcomes prior to estimation, causal machine learning approaches like causal forest use flexible data-driven forest-based constructions (\texttt{grf} also allow for more advanced use-cases by allowing for user-specified estimates of the first-stage inputs $e(x)$ and $m(x)$, which can be derived from separate machine learning models, such as gradient boosting or neural networks, or through ensemble methods that combine multiple models \citep{van2007super}).

\begin{rema} \label{rem:ehat}
In this context, an experimental setting is a \emph{special case} of an observational setting with a known propensity score. The implication is that the only algorithmic difference between analyzing experimental or observational data is that in the experimental setting, we supply the known propensity score instead of estimating it (in \texttt{causal\_forest}, this can be done via supplying the argument \texttt{W.hat}). The propensity score is then used the same way, including in downstream analyses, such as in calculating doubly robust average treatment effects discussed in Section \ref{sec:ate} (statistical guarantees on average treatment effect estimation are then even stronger, see \citet[Corollary 3.3]{wagerbook} for details).
\end{rema}

\section{Application: Estimating Resilience}\label{sec:application}
We illustrate the approach by focusing on a recent application of \texttt{grf} to estimate differential resilience to the effects of combat stress in leading to post-traumatic stress disorder (PTSD) among US Army soldiers \citep{kesslerPTSD}. PTSD is the signature mental disorder of war \citep{paulson2007haunted}. The roughly 7\% of the US population made up of military personnel and veterans are estimated to account for nearly 20\% of all cases of PTSD in the US, with an annual societal cost estimated at more than \$230B \citep{deangelis2023a}. However great variation is thought to exist in the effects of combat stress on PTSD among service members \citep{karstoft2015early, schultebraucks2021pre}.
We illustrate the value of causal forests by investigating these effects in a secondary analysis of a sample of US Army soldiers in three combat brigades who participated in a self-reported survey shortly before and then again after returning from a combat deployment in Afghanistan in 2012-2014 as part of the Army Study to Assess Risk and Resilience in Servicemembers (Army STARRS, \citet{ursano2014army}). Exposure to combat stress was assessed in the follow-up surveys administered to this sample after they returned from deployment. As expected, high combat stress exposure was found to be associated significantly with elevated prevalence of current PTSD as assessed in the follow-up surveys \citep{kesslerPTSD}.

The effects of exposure to high combat stress can be approximated by defining a dichotomous measure of the extent of combat stress exposure in the sample who had combat arms occupations (e.g., infantry, artillery). The extent to which information about individual differences in both baseline survey reports (including reports about history of and recent symptoms of PTSD) and baseline administrative data predicted this variation in exposure to high combat stress can then be investigated. The net association of high combat stress with the subsequent occurrence of meeting diagnostic criteria for PTSD in the follow-up survey can then be examined to estimate ATE as well as to inspect the distribution of estimated CATE.  More details on the dataset \citep{papini2023development} and a more extensive substantive analysis of the data \citep{kesslerPTSD} are presented elsewhere.

In order to use methods for CATE estimation to capture heterogeneity in resilience, we need to somewhat reconceptualize the notion of an ``intervention'': The intervention here, i.e., member combat deployment, is not a therapeutic intervention as is often considered in medical settings, but is nonetheless one that can be manipulated in meaningful ways. Only a fraction of all military personnel in a unit (often in the range 40-60\%) are assigned to deploy, with the remainder held in reserve (referred to by the military as the ``rear detachment''). Commanders take a wide range of factors into consideration in deciding which soldiers to deploy and which to hold in reserve. Estimated mental fitness based on supervisor assessment is one of these considerations. A CATE estimate based on a comprehensive assessment of administrative data and an optimized algorithm based on causal forests could be a useful addition to such assessments either to help commanders decide which members of their units to deploy and which to keep in the rear detachment or, in cases where it is necessary to deploy a non-resilient service member based on other considerations, to help target special resilience-enhancing resources (e.g., special pre-deployment resilience training programs \citep{thompson2018applicability}).

Considering deployment with few traumatic events (``low combat stress'') as a treatment arm ($W_i=1$) and deployment with ``high combat stress'' as a control arm ($W_i=0$), we can construct a causally motivated empirical measure of resilience using the language of potential outcomes. Table \ref{tab:AppliedPOs} shows an example of potential outcome configurations, or principal strata, describing outcomes a unit would experience in both treatment states. These principal strata are not observable (because, like the ITE, they depend on both potential outcomes); however, they are still useful for interpretation \citep{frangakis2002principal}.
The type of soldier who is healthy regardless of combat stress exposure can be deemed to have high resilience, whereas the type of soldier who develops PTSD under high combat stress but is healthy under low combat stress can be deemed to have low resilience. Finally, the type of soldier that develops PTSD regardless of combat stress exposure can be termed high risk. We assume that there are no soldiers who would be healthy under high stress and sick under low stress; this assumption is mathematically related to the widely used ``no defiers'' assumptions in clinical trials with non-compliance \citep{angrist1996identification}.

\begin{table}[t]
    \footnotesize
    \centering
    \begin{tabular}{ll|cc}
    & & \multicolumn{2}{c}{High combat stress}  \\
    & & Healthy & PTSD \\
    \hline
    \multirow{2}{*}{Low combat stress}
    & Healthy & ``High resilience''   & ``Low resilience''     \\
    & PTSD & $\times$  & ``High risk''     \\
    \end{tabular}
    \caption{Hypothetical potential outcomes configurations (``Healthy'', ``PTSD'') in the two treatment states (``low combat stress'' and ``high combat stress'') with corresponding resilience classifications.}
    \label{tab:AppliedPOs}
\end{table}

We record realized healthy outcomes by 1 and PTSD by 0,
$$
Y_i = 
\begin{cases}
    1 & \text{if Healthy} \\
    0 & \text{if PTSD},
\end{cases}
$$
and so our average treatment effect measures the fraction of healthy soldiers when no one is deployed minus the counterfactual fraction of healthy soldiers when all soldiers are deployed, which we expect to be larger than zero.
\begin{align*}\label{eqn:ate_def}
\begin{split}
    \tau &= \EE{Y_i(1) - Y_i(0)} \\
    &=\PP{Y_i(\text{Low combat stress})=\text{Healthy}} - \PP{Y_i(\text{High combat stress})=\text{Healthy}} \\
    &=\PP{\text{Low resilience}}.
\end{split}
\end{align*}
The CATE then captures heterogeneity in resilience:
$$
\tau(x) = \PP{\text{Low resilience} \mid X_i = x}.
$$
This type of taxonomy gives us two approaches to measuring resilience: a causal measure based on CATEs (low vs. high resilience) and a predictive measure based on risk (high vs. low risk). How these two approaches differ in practice is an empirical question we address in Section \ref{sec:riskbased}.

\paragraph{Illustration code.}
Illustration code for the \texttt{R} language \citep{Rcore} to conduct the analysis in this section with synthetic example data, using the package \texttt{grf} \citep{GRF} along with \texttt{maq} \citep{MAQ} and \texttt{ggplot2} \citep{ggplot}, is available on GitHub at \href{https://github.com/grf-labs/grf/tree/master/experiments/ijmpr}{github.com/grf-labs/grf/tree/master/experiments/ijmpr}.

\subsection{An Exploratory Analysis}\label{sec:analysis}
We consider 2,542 US soldiers with combat service, where 1,542 experienced low combat stress, and 1,000 soldiers experienced high combat stress (for details on the sample construction, including details on treatment and outcome definitions, we refer to \citet{kesslerPTSD}). In the high combat stress group, 96 soldiers developed PTSD post-deployment, while in the low combat stress group, 36 soldiers developed PTSD.
The simple difference-in-means (i.e., the mean number of healthy soldiers in the low-stress group minus the mean number of healthy soldiers in the high-stress group) is 7.4\% (standard error of 0.9\%), and yields a measure of the raw association between deployment and PTSD. This is, however, not necessarily a valid estimate of the ATE, as combat stress exposure during deployment might not be completely random.

In order to mitigate confounding, we control for a set of pre-treatment characteristics that includes variables describing individual disorders, suicidality measures, past trauma experiences, past stressors, medical treatments, personality scores, army-specific employment details, demographics, and details on traumatic brain injuries, for a total of 410 variables. If we are willing to maintain an unconfoundedness assumption, then we can use these variables to account for the different proportions of soldiers with, for example, past trauma experiences in the two combat groups. Using a causal forest, a doubly robust estimate that adjusts for these pre-treatment characteristics gives an average treatment effect of 6.3\% (standard error of 1\%). \texttt{grf} forms this estimate via forest-based Augmented Inverse-Propensity Weighting (AIPW, \citet{robins1994estimation}) using the causal forest first-stage estimates of the propensity score $e(x)$ and conditional mean $m(x)$.

\begin{figure}[ht]
\centering
    \includegraphics[width=0.6\textwidth]{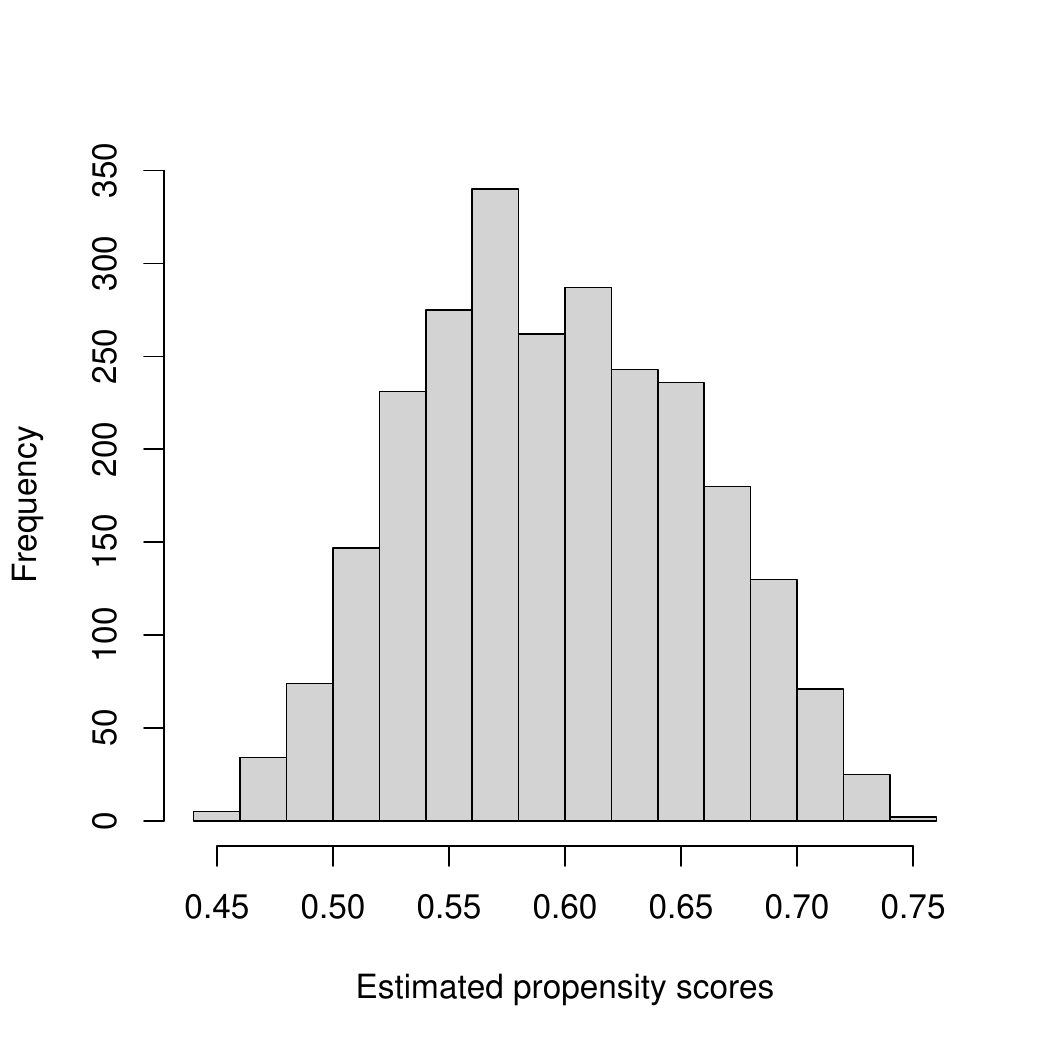}
\caption{Histogram of propensity scores estimated with a \texttt{grf} regression forest.}
\label{fig:ehat}
\end{figure}
A necessary condition for this to be a viable strategy for estimating causal effects is that the propensity score as a function of these variables is bounded away from zero and one, i.e., we have treatment overlap. Figure \ref{fig:ehat} shows a histogram of the forest-estimated propensity scores and indicates the overlap assumption is plausible as most of the propensity scores fall within the range 0.5-0.7, that is, along our observable adjustment variables the different soldiers are more or less equally likely to be in either treatment group.

\subsection{Treatment Effect Heterogeneity}\label{sec:cateforest}

The average-case analysis above established that there are a substantial number of low-resilience soldiers in the sample.
However, there might be between-individual variation in the effects of high combat stress such that prevalence of PTSD could be reduced by taking this individual variation into account. In order to take a data-driven approach to discover potential subgroups of soldiers that respond differently to combat stress, we divide the dataset into two random groups. In the training set (60\% of the data), we fit a CATE estimator, and on a held-out test set (40\% of the data), we evaluate these CATE estimates (the exact train/test ratio is more or less a heuristic; we prefer to have slightly more data for training than evaluation in this small dataset). 

\begin{figure}[ht]
    \centering
    \begin{subfigure}[b]{0.5\textwidth}
        \centering
        \includegraphics[width=\textwidth]{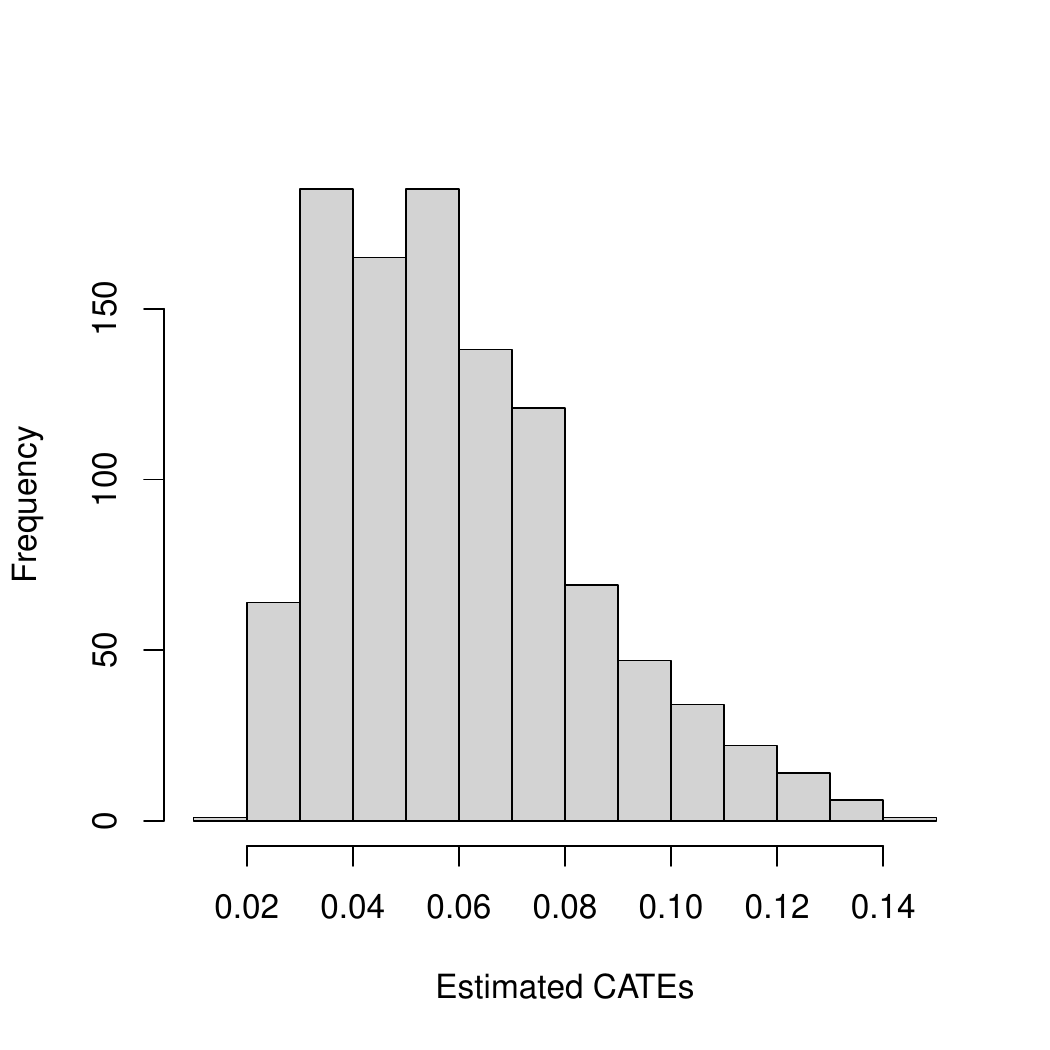}
        \caption[]
        {Histogram of causal forest CATE point estimates, $\hat \tau(X_i)$, for soldiers on the held-out test set.\newline}
        \label{fig:cate_hist}
    \end{subfigure}
    \hfill
    \begin{subfigure}[b]{0.475\textwidth}  
        \centering 
        \includegraphics[width=\textwidth]{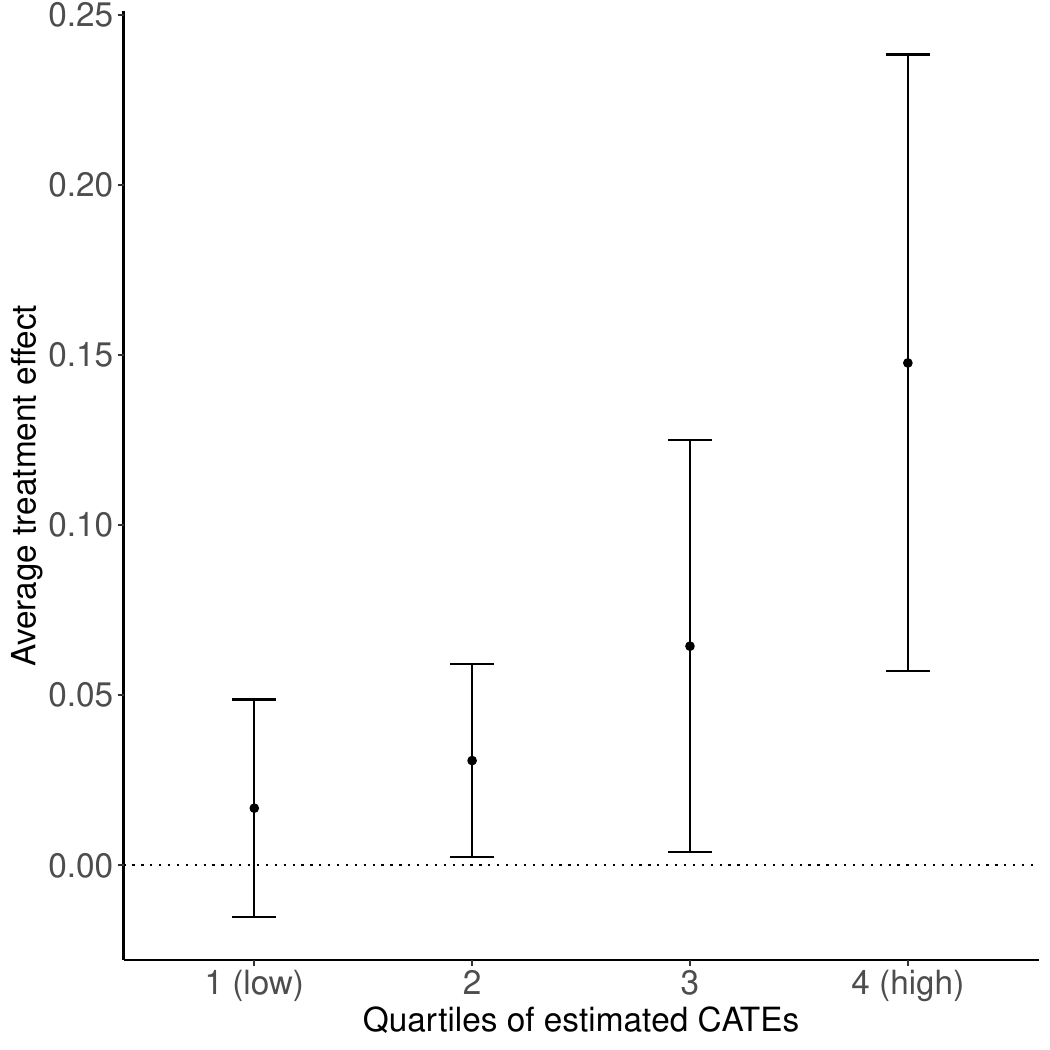}
        \caption[]
        {Average treatment effects by quartiles of the test set CATE estimates, along with 95\% confidence bars. The average treatment effect in each quartile is estimated with forest-based AIPW.}    
        \label{fig:quartiles}
    \end{subfigure}
    \caption{Illustration of the trained CATE function (histogram of predictions in (a)) capturing meaningful treatment heterogeneity in test set subgroups (b).}\label{fig:cate_hist_quartiles}
\end{figure}

On the training data, we fit a causal forest using default settings for hyper-parameters and obtain an estimated CATE function $\hat \tau(\cdot)$. These default settings often work well, though with \texttt{grf} it is possible to select parameters with automated tuning that minimize the \emph{R}-loss criterion of \citet{nie2020quasi}. Figure \ref{fig:cate_hist} shows a histogram of the CATE estimates for units on the test sample. The figure suggests that there are soldiers who would benefit meaningfully more than average from not experiencing high combat stress. There also appear to be other soldiers that would be much less affected than average from experiencing high combat stress. However, this histogram is made up of individual point estimates that are inherently noisy. To assess the statistical significance of this variation in CATE estimates, we need to use our test set. We do this by using the model developed in the training sample to predict CATEs in the independent test set. We then group soldiers in the test set sample into buckets according to which quartile of the predicted CATEs they belong to. Finally, we estimate average treatment effects in each bucket via AIPW (using a separate test set causal forest). Figure \ref{fig:quartiles} shows an estimate of the average treatment effect over these quartiles. The results suggest that treatment effect heterogeneity is statistically significant and substantively robust. In the quartile of the sample with the lowest predicted CATEs the average treatment effect is close to zero, indicating that exposure to high combat stress would be expected to have no significant effect in increasing risk of PTSD. In the quartile with the highest predicted CATEs, in comparison, the average treatment effect is large and statistically significant, indicating that exposure to high combat stress would have a substantial effect in increasing risk of PTSD. The estimated average treatment effects in the two intermediate quartiles fall between the extremes found in the low and high quartiles.

\paragraph{A $p$-value for heterogeneity.}
It is noteworthy that we chose to divide the test sample into quartiles because it seemed reasonable given the relatively small sample size, although other perfectly reasonable choices would have been tertiles, quintiles, or, in larger samples, perhaps deciles. The Targeting Operator Characteristic curve (TOC) \citep{yadlowsky2021evaluating} provides an agnostic version of the visualization in Figure \ref{fig:quartiles} that does not require pre-specifying the number of groups. This is achieved by ranking soldiers according to predicted treatment effects and computing the average treatment effect for each $q$-th quantile in this ranking. This can then be compared to the overall average treatment effect in the following way:
$$
\text{TOC}(q) = \EE{Y_i(1) - Y_i(0) \mid \text{Estimated CATE}(X_i) \geq 1 - q\text{-th quantile}} - \text{ATE}.
$$
As we are subtracting the overall ATE, this curve will end at 0 for $q=1$. 

\begin{figure}[ht]
\centering
    \includegraphics[width=0.6\textwidth]{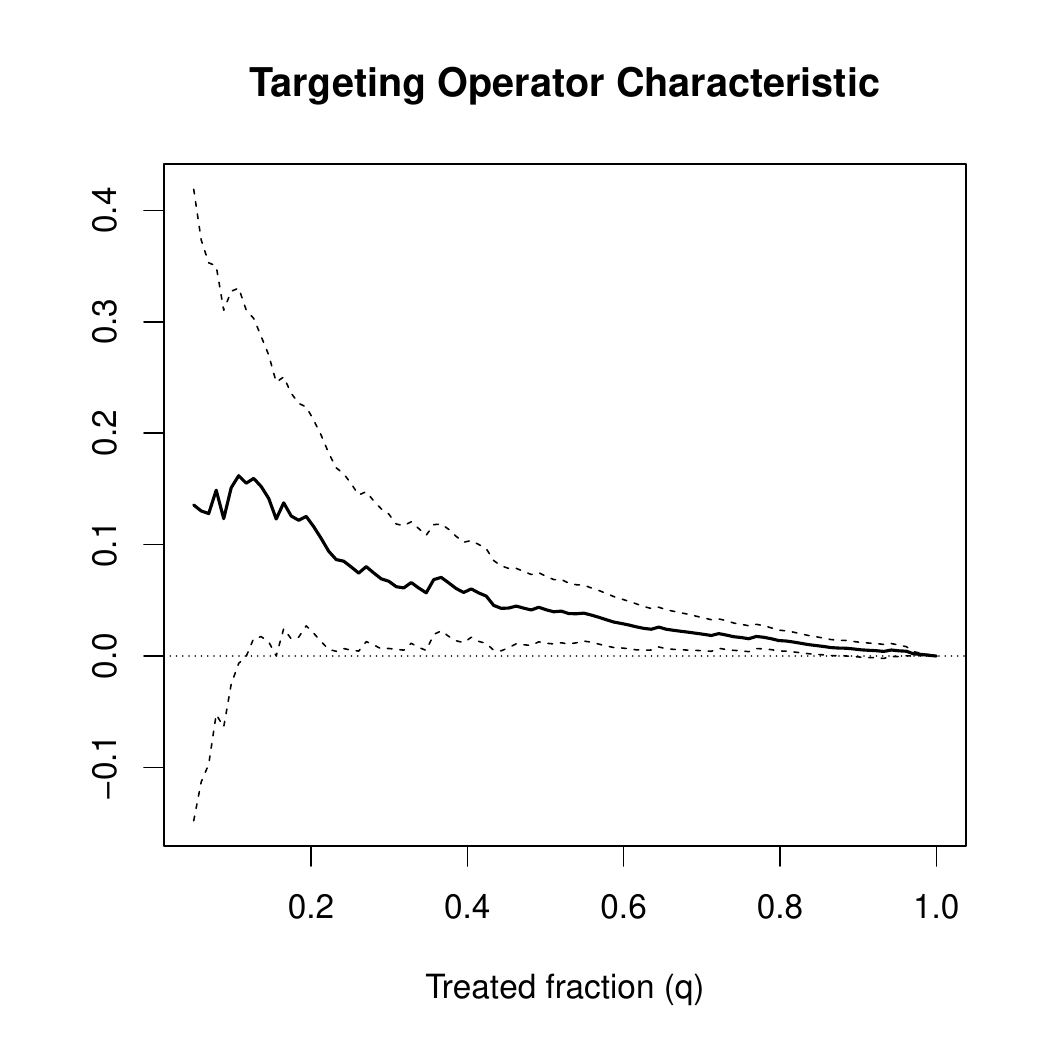}
\caption{Targeting Operator Characteristic curve for causal forest CATE predictions, estimated on a held-out test set. The dashed lines are pointwise 95\% confidence intervals that are conditional on our estimated CATE function and quantify test set uncertainty in estimating the TOC.}
\label{fig:rate_cate}
\end{figure}
Figure \ref{fig:rate_cate} shows the estimated TOC curve for the training set predicted CATEs, and indicates that there are signs of treatment effect heterogeneity. For example, the top $q=$ 20\% of soldiers with the largest predicted treatment effects have an ATE that is around 11\% higher than the overall ATE (with a standard error of $\approx$ 5\%). That is, the group of soldiers belonging to the top quintile of predicted treatment effects had low resilience at an 11\% higher rate than average. The confidence intervals are wider at lower quantiles than at higher quantiles since the number of observations is smaller. In this setting, the lowest quantiles of the TOC curve cover zero, highlighting there is considerable estimation uncertainty for this smaller group. We can translate the visual stratification exercise underlying the TOC curve into a single-point estimate by forming an estimate of the area under this curve. 

An estimate of the area under the TOC (``AUTOC'') is 0.063 with a bootstrapped standard error of 0.028. The AUTOC satisfies a central-limit theorem \citep{yadlowsky2021evaluating}, so we can motivate a test for heterogeneity with
\begin{align*}
&H_0: \text{AUTOC} = 0 ~~~ \text{(No heterogeneity detected by estimated CATE function)}\\
&H_A: \text{AUTOC} \neq 0,
\end{align*}
using the $t$-value ${\widehat{\text{AUTOC}}}\,/\,{\sqrt{\text{Var}[\widehat{\text{AUTOC}}}]} = \frac{0.063}{0.028}=2.25$. An asymptotically valid $p$-value for this two-sided test is $2\PP{z > |2.25|} = 0.02$ (where $z$ is a standard random normal). We reject $H_0$ at a significance level of $\alpha = 5\%$, which suggests that we have successfully identified non-zero treatment heterogeneity using our CATE estimate.

Beyond serving as a valuable test for heterogeneity, the AUTOC can also assist in model selection. For instance, when comparing candidate predictor models (e.g., two or more alternative CATE models), the model with the highest AUTOC is the one that most effectively stratifies the test set sample. As mentioned in Section \ref{sec:CF}, causal forests are a two-step algorithm where the first step accounts for confounding with respect to some specified baseline predictor variables via an estimated propensity score as well as baseline effects via conditional mean estimates and the second step fits a CATE function based on some set of baseline predictor variables. In this application, compared to our sample size, the set of predictor variables considered was large (410), so we began the analysis by using a simple pre-screening heuristic to reduce the number of variables to help our forest capture relevant heterogeneity \citep{athey2019estimating}: For the CATE model, we only used predictor variables that, in the conditional mean model, had variable importance in the top 25-th percentile. The AUTOC for this restricted CATE model was very similar to that for the full predictor set, indicating that this subset of predictors captured parsimoniously the meaningful heterogeneity associated with the full predictor set; this approach is closely related to evaluating binary classifiers with the ROC curve in a predictive analysis. Section \ref{sec:riskbased} gives one more example of using the AUTOC for evaluating treatment targeting strategies. 
\begin{rema} \label{rem:varimp}
\texttt{grf} uses a simple assessment of variable importance, first discussed by \citet{breiman2001random}, that counts how often a particular variable is used for recursive partitioning throughout the forest. For a broader discussion of alternative variable importance measures with forests see \citet{benard2023variable} and \citet{hastie2009elements}.
\end{rema}

\subsection{Evaluating treatment rules with Qini curves}
The exercise and plots in the preceding section can help visualize and get a single-point estimate for evidence of heterogeneity captured by our estimated CATE function. There are numerous use cases for this function. As noted above, one important case would be to use this type of formulation to help decide which soldiers either to hold back from deployment or to provide additional resilience training prior to deployment. A treatment allocation policy of this sort could take many forms. It could, for example, target some fixed proportion of soldiers with the largest predicted effects and then look at what overall reduction in PTSD might be expected. The Qini curve is a simple visualization that plots the overall expected treatment effect on this aggregate PTSD rate obtained by intervening in this way over all possible thresholds of number of soldiers.
\begin{figure}[ht]
\centering
    \includegraphics[width=0.6\textwidth]{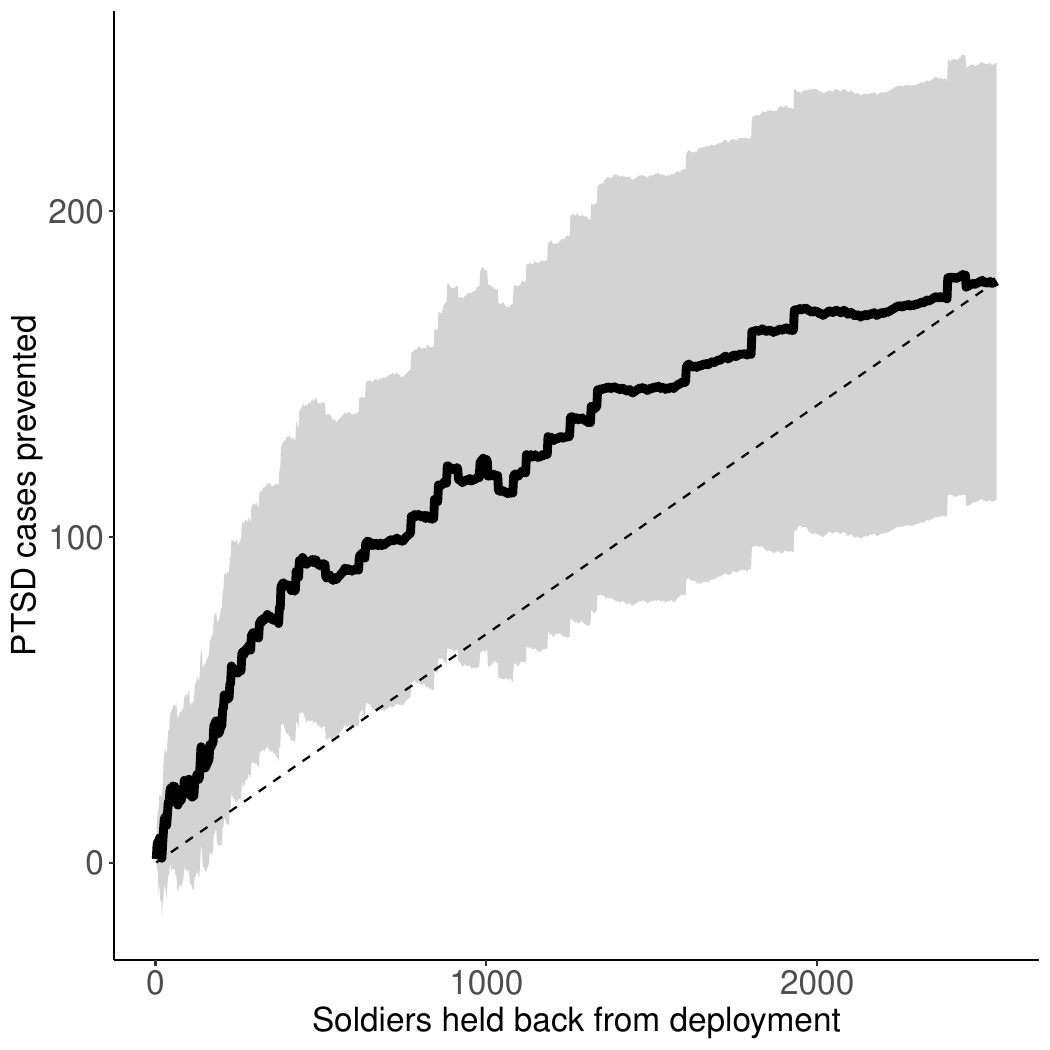}
\caption{Qini curve for the policy that uses our test set predicted CATEs to hold back soldiers (solid black line, 95\% interval in shaded region). The dashed straight line is a Qini curve for a policy that holds back an arbitrary group of soldiers and traces out an average treatment effect. The confidence intervals are conditional on the estimated CATE function and reflect estimation uncertainty from deploying a policy on a random test set.}
\label{fig:qini}
\end{figure}

Figure \ref{fig:qini} shows the Qini curve for this case study based on our estimated CATE function for our sample of $n=$ 2,542 soldiers. The straight dashed line in Figure \ref{fig:qini} is a non-targeting baseline policy that shows the reduction in total PTSD cases we could expect by intervening randomly with various numbers of soldiers. The far right of this curve equals $n \cdot \widehat{\text{ATE}} \approx 180$ soldiers, the PTSD cases avoided if no one is deployed. The solid black curve shows results that can be achieved by using our CATE estimate to target intervention. For example, if we use causal forests to choose 500 soldiers predicted to benefit most, we estimate an overall reduction in PTSD cases of around 90 cases (standard error of 25). To achieve the same benefit by intervening randomly, we would have to hold back 1,300 soldiers.

\subsection{Describing estimated heterogeneity via covariate profiles}\label{sec:cate_profile}
A reasonable question to ask based on the material described thus far is how soldiers differ along observable pre-treatment characteristics in groups that benefit more vs. less from treatment. A deeper question underlying this initial question is whether meaningful psychiatric risk profiles emerge from such an investigation. Such questions can be addressed by zooming in on individuals in the lower and higher quartiles of Figure \ref{fig:quartiles} to look at the distribution of some candidate predictor variables.  Given that, as noted above, the predictor set is relatively large, it is convenient to begin by inspecting the variable importance using the causal forest variable importance measure described above (had we not had access to this particular variable importance measure, perhaps because we used another method than causal forest, then another simple heuristic to narrow down which variables to look at could be, for example, to choose the ones where the standardized differences in means over the high/low subgroups are more than, say, one standard deviation). When this was done, we found that a handful of baseline 30-day symptom measures had very high importance rankings. As an exploratory step, we look at the top-6 such variables.
\begin{figure}[t]
\centering
    \includegraphics[width=0.75\textwidth]{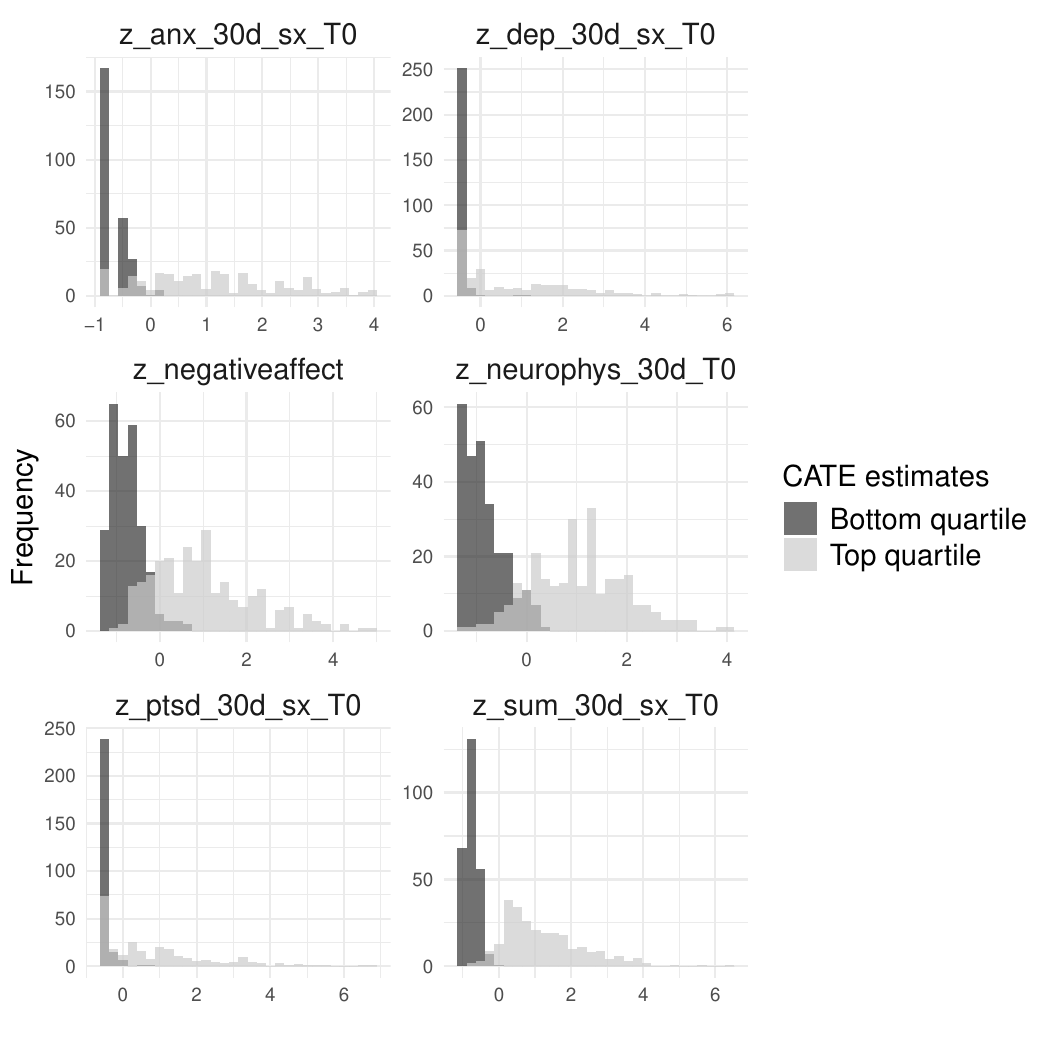}
\caption{Histograms of 6 predictor variables for soldiers that are in the bottom (resilient) and top quartile (less resilient) of predicted test set CATEs (Figure \ref{fig:quartiles}). These 6 predictor variables have high variable importance as measured by the causal forest split frequency. The variables are: ``z\_anx\_30d\_sx\_T0'' (anxiety symptoms), ``z\_dep\_30d\_sx\_T0'' (depression symptoms), ``z\_negativeaffect'' (negative affectivity), ``z\_neurophys\_30d\_T0'' (neuro/physiological symptoms), ``z\_ptsd\_30d\_sx\_T0'' (PTSD symptoms), and ``z\_sum\_30d\_sx\_T0'' (all symptoms). All variables are constructed as a standardized continuous score for 30-day symptoms.}
\label{fig:variable_histogram}
\end{figure}

Figure \ref{fig:variable_histogram} shows a histogram of these variables for the groups of soldiers that belong to either the test set bottom or top quartile of predicted CATEs. We can see clear evidence that the distribution differs. This handful of variables captures an assortment of health measures (anxiety, PTSD symptoms, neurosis, etc.) and indicates that soldiers that benefit from being held back from deployment score higher on these indices (i.e, they have more severe 30-day health measures than the soldiers in the bottom quartile). At a high level, these covariate differences confirm an intuition that soldiers at highest risk of PTSD already had poor mental health prior to deployment.

\subsection{Inference on effect modifiers via best linear projections}
The exploratory heterogeneity exercise in Section \ref{sec:cate_profile} can be very useful to get an insight into how our estimated CATE function ended up stratifying our data in terms of various covariate profiles. In order to relate candidate predictor variables to exact statistical quantities, such as average effects, we need other approaches. One appealing approach is via a linear summary measure of the CATE. The best linear projection is the optimal prediction of CATE as a function of some chosen variable(s) $Z_i$,
\begin{equation}\label{eqn:blp_def}
    \tau(X_i) \sim \alpha + \beta Z_i + \varepsilon.
\end{equation}
Formally, we have $\{\alpha^*, \, \beta^*\}  = \argmin_{\alpha, \beta}\left\{ \EE{(\tau(X_i) - \alpha - Z_i \beta)^2}\right\}$. This is not a standard linear regression problem as $\tau(X_i)$ is not observable. However, \texttt{grf} enables inference about $\beta$ via a built-in AIPW-style construction that has similar statistical properties as what one would expect of classical parametric summaries and can be used to justify exact confidence intervals \citep{semenova2021debiased}. In forming estimates like these, typical considerations that arise in the familiar parametric setting apply concerning multiple testing and pre-specified hypotheses. Note that, when we deploy a best linear projection we are \emph{not} assuming that the data is linear, we are simply summarising a non-linear object, the CATE, via a linear approximation.
\begin{table}[ht]
\footnotesize
\centering
\begin{tabular}{l c c c c c}
\hline
 & (1) & (2) & (3) & (4) & (5) \\
 
\hline
(Intercept)           & $0.06^{***}$ & $0.06^{***}$ & $0.06^{***}$ & $0.06^{***}$ & $0.06^{***}$ \\
                      & $(0.01)$     & $(0.01)$     & $(0.01)$     & $(0.01)$     & $(0.01)$     \\
z\_anx\_30d\_sx\_T0   & $0.05^{***}$ &              &              &              & $0.06$       \\
                      & $(0.02)$     &              &              &              & $(0.03)$     \\
z\_neurophys\_30d\_T0 &              & $0.04^{***}$ &              &              & $0.04$       \\
                      &              & $(0.01)$     &              &              & $(0.03)$     \\
z\_ptsd\_30d\_sx\_T0  &              &              & $0.04^{*}$   &              & $0.03$       \\
                      &              &              & $(0.02)$     &              & $(0.03)$     \\
z\_sum\_30d\_sx\_T0   &              &              &              & $0.05^{**}$  & $-0.06$      \\
                      &              &              &              & $(0.02)$     & $(0.06)$     \\
\hline
\multicolumn{6}{l}{\scriptsize{$^{***}p<0.001$; $^{**}p<0.01$; $^{*}p<0.05$}}
\end{tabular}

\caption{Estimates of best linear projections \eqref{eqn:blp_def} using \texttt{grf} for 4 predictors: ``z\_anx\_30d\_sx\_T0'' (anxiety symptoms), ``z\_neurophys\_30d\_T0'' (neuro/physiological symptoms), ``z\_ptsd\_30d\_sx\_T0'' (PTSD symptoms), and ``z\_sum\_30d\_sx\_T0'' (all symptoms). All variables are constructed as a standardized continuous score for 30-day symptoms. $HC_3$ \citep{mackinnon1985some} standard errors in parentheses.}
\label{table:blp}
\end{table}

We consider 4 potential effect modifiers $Z_i$. Table \ref{table:blp} shows estimates of best linear projections first via separate linear models, then finally jointly in the last column. We can interpret the coefficients in this table just as we would with any other regression model. Recall we defined our desirable outcome to be 1 if healthy and the causal effects we defined in Section \ref{sec:application} is measuring a counterfactual difference in PTSD prevalence. We can see that a one-unit increase in ``z\_anx\_30d\_sx\_T0'' (anxiety symptoms score) is related to a 5\% increase in PTSD,  and similarly for the other variables. These health indicators are all likely to be correlated, as indicated by the last column in Table \ref{table:blp}. Table \ref{table:corr} shows the empirical correlation matrix of these variables.
\begin{table}[ht]
\footnotesize
\centering

\begin{tabular}{rrrrr}
  \hline
 & z\_anx\_30d\_sx\_T0 & z\_neurophys\_30d\_T0 & z\_ptsd\_30d\_sx\_T0 & z\_sum\_30d\_sx\_T0 \\ 
  \hline
z\_anx\_30d\_sx\_T0 & 1.00 & 0.56 & 0.48 & 0.85 \\ 
  z\_neurophys\_30d\_T0 & 0.56 & 1.00 & 0.42 & 0.78 \\ 
  z\_ptsd\_30d\_sx\_T0 & 0.48 & 0.42 & 1.00 & 0.72 \\ 
  z\_sum\_30d\_sx\_T0 & 0.85 & 0.78 & 0.72 & 1.00 \\ 
   \hline
\end{tabular}

\caption{Empirical correlation table for the predictor variables described in Table \ref{table:blp}.}
\label{table:corr}
\end{table}

\subsection{Risk-based targeting and treatment rule model selection}\label{sec:riskbased}
Recall that in our conceptual resilience taxonomy in Table \ref{tab:AppliedPOs}, we labeled soldiers that experienced PTSD regardless of treatment state as ``high risk''. In our empirical setting, it seems plausible that resilience and risk-based classifications are overlapping categories. As we've seen in Figure \ref{fig:variable_histogram}, the type of people more affected by high combat stress appear to have high baseline prevalence of psychiatric problems. One reasonable risk model for our setting could be the following,
$$
\text{risk}(X_i) = \EE{\mathbf{1}(Y_i = \text{PTSD}) \mid X_i,~ W_i = \text{high combat stress}},
$$
which is the probability of a soldier with covariates $X_i$ developing PTSD when exposed to high combat stress. Using \texttt{grf} we can form a forest-based estimate of this model by fitting a regression forest on the subset of training set soldiers exposed to high combat stress.

\begin{figure}[t]
    \centering
    \begin{subfigure}[b]{0.485\textwidth}
        \centering
        \includegraphics[width=\textwidth]{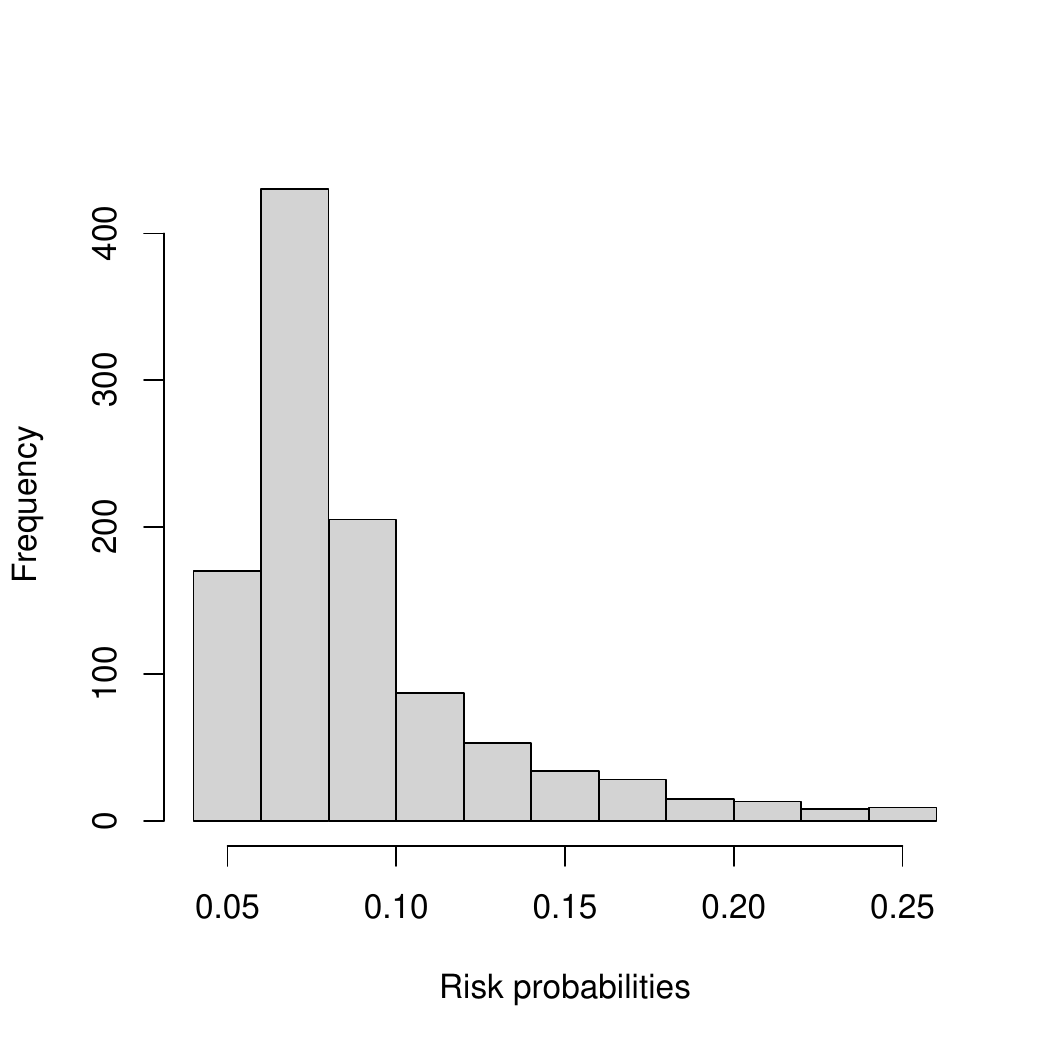}
        \caption[]
        {Histogram of estimated test set risk probabilities, $\widehat{\text{risk}}(X_i)$, using a regression forest fit on high combat stress soldiers on the training set.}
        \label{fig:risk_hist}
    \end{subfigure}
    \hfill
    \begin{subfigure}[b]{0.485\textwidth}  
        \centering 
        \includegraphics[width=\textwidth]{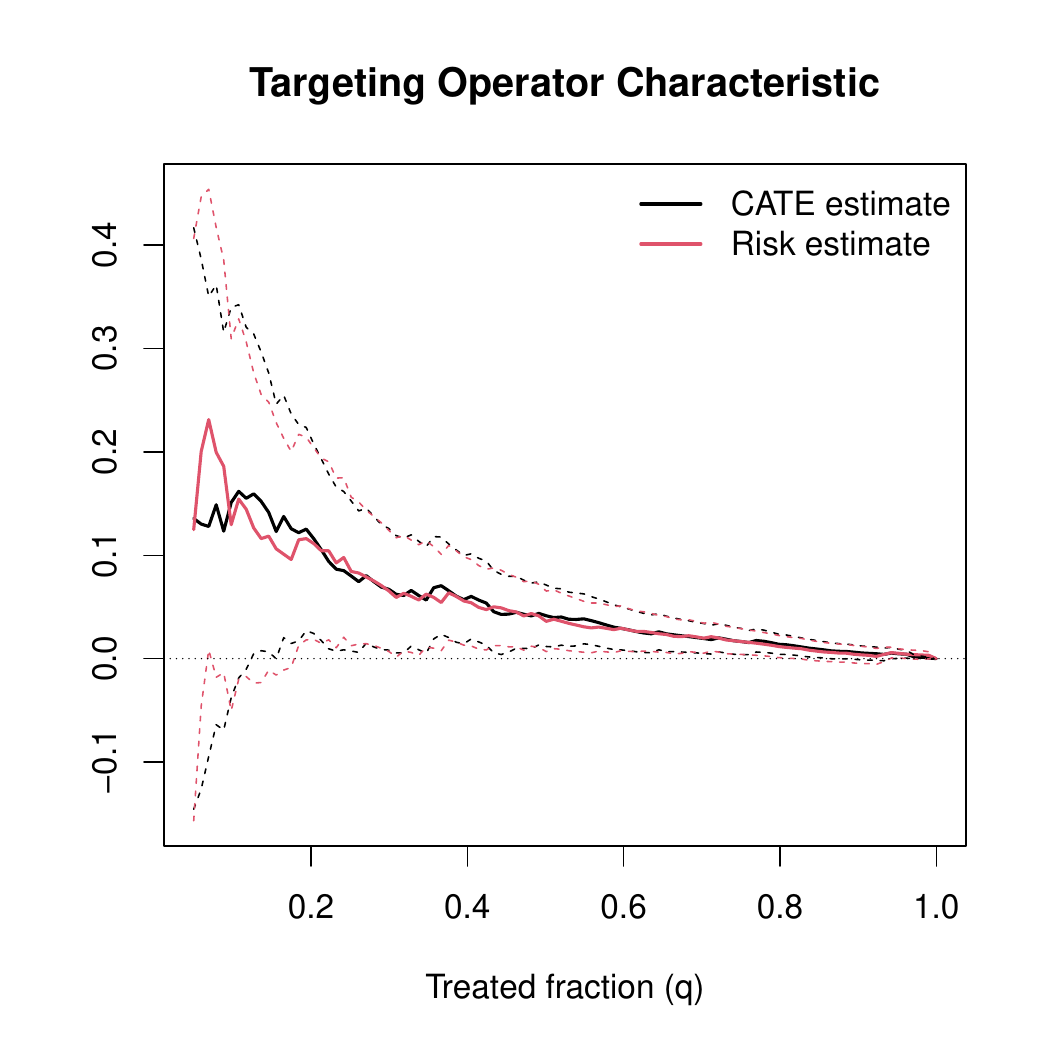}
        \caption[]
        {TOC curve comparing prioritizing treatment with estimated CATEs and with estimated risk.\\}    
        \label{fig:rate_risk}
    \end{subfigure}
    \caption{Comparing predictive (risk-based) and causal (CATE) targeting.}\label{fig:risk}
\end{figure}

Figure \ref{fig:risk_hist} shows a histogram of this estimated probability for soldiers in the held-out test sample (where we used all 410 pre-treatment characteristics as predictor variables). The question is, do these risk estimates perform well in targeting soldiers with different treatment benefits? Our intuition is that soldiers that have a large predicted risk of developing PTSD, also have a large treatment benefit of intervention. Using the same quartile stratification exercise as in Figure \ref{fig:quartiles}, we could stratify the sample by risk quartiles and then estimate average treatment effects. However, while visually appealing, our end goal is to compare targeting strategies, and comparing quartile figures is not as convenient for this purpose. To this end, it turns out we can use the TOC construction to facilitate a side-by-side comparison.

Recall the TOC curve we used to evaluate our estimated CATEs is defined by the following
$$
\text{TOC}(q) = \EE{Y_i(1) - Y_i(0) \mid \text{Estimated CATE}(X_i) \geq 1 - q\text{-th quantile} } - \text{ATE},
$$
where key point to recognize is that the relative ranking of the estimated CATEs makes up this curve. This means we can use this same construction to evaluate other intervention strategies as long as they imply a ranking/prioritization. Using our estimated risk model to prioritize soldiers with high  probabilities of PTSD, we get the TOC:
$$
\text{TOC}_{\text{risk}}(q) = \EE{Y_i(1) - Y_i(0) \mid \text{Estimated risk}(X_i) \geq 1 - q\text{-th quantile} } - \text{ATE}.
$$
Figure \ref{fig:rate_risk} compares the TOC curve using risk predictions with a TOC curve using our estimated CATEs and reveals that they perform very similarly. The area under the curve using risk-based predictions is close to the CATE-based prediction: 0.065 vs 0.063. We can construct a test for whether these two targeting strategies perform the same by estimating the difference in the area under the curve and calculating a $p$-value (or equivalently, an $\alpha$-level confidence interval). The resulting $p$-value is 0.8 and suggests that these two targeting strategies are statistically indistinguishable. The implication is that for our intended application, risk-based targeting may be good enough.

Finally, as another example of the use case for the TOC curve; recall that Figure \ref{fig:variable_histogram} reveals that a handful of 30-day health symptoms scores appear to differ markedly for soldiers with different treatment benefits. These variables are all continuous-valued, and as noted in Table \ref{table:corr} highly correlated with each other. We can use the TOC to test if a simple treatment prioritization using one of these variables does well in classifying soldiers on differential effects of high combat exposure on PTSD. If we pick the variable ``standardized sum of 30-day health symptoms'' we get a TOC that focuses only on that single predictor. Doing so, we get an AUTOC that is similar to both the risk-based and CATE-based ones: 0.068 (standard error 0.024), and that is not distinguishable from either if we compare the difference in AUTOCs.

\begin{rema} \label{rem:risk}
We estimated a risk model directly using a random forest with all available predictor variables. In some cases, it may be possible to get better performance by first pre-screening pre-treatment characteristics using general variable selection tools, such as the Lasso. For a practical guide to developing risk models in clinical applications, see \citet{steyerberg2009applications}. 
\end{rema}

\section{Discussion}
We have surveyed how causal forests can be used to generate both principled estimates of ATE in observational studies and to detect treatment heterogeneity in experimental or observational studies. For more details on \texttt{grf}, including additional functionality such as missing values support, tree-based policy learning, loss-to-follow-up corrections, and more, we refer to vignettes and journal references on the \texttt{grf} website at \href{https://grf-labs.github.io/grf/}{grf-labs.github.io/grf}.

We have also outlined a general framework for analyzing and comparing output from causal machine learning algorithms on a level playing field with tools such as TOC and Qini curves. As noted in the introduction, there are a number of alternative machine learning approaches for CATE estimation that are popular in practice \citep[e.g.,][]{athey2016recursive, hahn2020bayesian, kennedy2023towards, kunzel2019metalearners, luedtke2016super, nie2020quasi, tian2014simple}. Customizable \texttt{Python} libraries that implement several of these methods include \texttt{causalML} \citep{chen2020causalml} and \texttt{econML} \citep{econml}; and the software toolkit for machine learning-based causal inference is still rapidly growing. In practice, it can sometimes be a good idea to try multiple different machine learning-based approaches to CATE estimations; the TOC, Qini curves, and related tools can then be used to compare and benchmark their performance.

\clearpage
\section*{Acknowledgments}
We are grateful to the STARRS-LS collaborators for the use of their survey data and to Ron Kessler for providing invaluable feedback and advice.\\ 

\noindent \textbf{Author Contributions}: Petukhova had full access to all study data and takes responsibility for the integrity of the data and the accuracy of the data analysis.\\

\noindent \emph{Concept and design}: Sverdrup, Wager.\\

\noindent \emph{Acquisition, analysis, or interpretation of data}: All authors.\\

\noindent \emph{Drafting of the manuscript}: Sverdrup, Petukhova.\\

\noindent \emph{Critical revision of the manuscript for important intellectual content}: All authors.\\

\noindent \emph{Statistical analysis}: Petukhova.\\

\noindent \emph{Obtained funding}: NA.\\

\noindent \emph{Administrative, technical, or material support}: Sverdrup.\\

\noindent \textbf{Conflict of Interest Disclosures}:
No disclosures were reported.\\

\noindent \textbf{Conflict of Interest Disclosures}:
The study protocol was approved by the Research Ethics Committees of the Henry Jackson Foundation and Harvard Medical School (IRB15-0765) with a waiver of informed consent based on data being de-identified. Research has been performed in accordance with the Declaration of Helsinki.\\

\noindent \textbf{Funding/Sponsor}: Army STARRS was sponsored by the Department of the Army and funded under cooperative agreement number U01MH087981 (2009-2015) with the U.S. Department of Health and Human Services, National Institutes of Health, National Institute of Mental Health (NIH/NIMH). Petukhova has subsequently been supported by STARRS-LS, which is sponsored and funded by the Department of Defense (USUHS grant number HU0001-15-2-0004). This support funded her participation in this paper. The contents are solely the responsibility of the authors and do not necessarily represent the views of the Department of Health and Human Services, NIMH, or the Department of the Army, or the Department of Defense. The dataset was made available by the Army STARRS Team, which consists of Co-Principal Investigators: Robert J. Ursano, MD (Uniformed Services University of the Health Sciences) and Murray B. Stein, MD, MPH (University of California San Diego and VA San Diego Healthcare System). Site Principal Investigators: Steven Heeringa, PhD (University of Michigan), James Wagner, PhD (University of Michigan) and Ronald C. Kessler, PhD (Harvard Medical School); Army scientific consultant /liaison: Kenneth Cox, MD, MPH (Office of the Deputy Under Secretary of the Army). Other team members: Pablo A. Aliaga, MA (Uniformed Services University of the Health Sciences); COL David M. Benedek, MD (Uniformed Services University of the Health Sciences); Laura Campbell-Sills, PhD (University of California San Diego); Carol S. Fullerton, PhD (Uniformed Services University of the Health Sciences); Nancy Gebler, MA (University of Michigan); Robert K. Gifford, PhD (Uniformed Services University of the Health Sciences); Meredith House, BA (University of Michigan); Paul E. Hurwitz, MPH (Uniformed Services University of the Health Sciences); Sonia Jain, PhD (University of California San Diego); Tzu-Cheg Kao, PhD (Uniformed Services University of the Health Sciences); Lisa Lewandowski-Romps, PhD (University of Michigan); Holly Herberman Mash, PhD (Uniformed Services University of the Health Sciences); James A. Naifeh, PhD (Uniformed Services University of the Health Sciences); Tsz Hin Hinz Ng, MPH (Uniformed Services University of the Health Sciences); Matthew K. Nock, PhD (Harvard University); Nancy A. Sampson, BA (Harvard Medical School); COL Gary H. Wynn, MD (Uniformed Services University of the Health Sciences); and Alan M. Zaslavsky, PhD (Harvard Medical School).\\

\noindent \textbf{Role of Funder/Sponsor}: As a cooperative agreement, scientists employed by the National Institute of Mental Health and U.S. Army liaisons and consultants collaborated to develop the STARRS study protocol and data collection instruments, supervise data collection, interpret results, and prepare reports. Although a draft of the manuscript was submitted to the U.S. Army and National Institute of Mental Health for review and comment before submission for publication, this was done with the understanding that comments would be no more than advisory.\\

\noindent \textbf{Data availability}:
Illustration code for the \texttt{R} language \citep{Rcore} to conduct the analysis with synthetic example data is available on GitHub at \href{https://github.com/grf-labs/grf/tree/master/experiments/ijmpr}{github.com/grf-labs/grf/tree/master/experiments/ijmpr}.

\clearpage
%IJMPR: apa bib
\bibliographystyle{apalike}
\bibliography{bibliography}

\clearpage
\newpage
\appendix

\end{document}